# Controllable Spatial Array of Bessel-like Beams with Independent Axial Intensity Distributions for Laser Microprocessing

Sergej Orlov, Alfonsas Juršėnas, Justas Baltrukonis and Vytautas Jukna

*Center for Physical Sciences and Technology, Industrial laboratory for photonic technologies, Saulėtekio av. 3, Vilnius, Lithuania*
*E-mail: Sergejus.orlovas@ftmc.lt*

Bessel beams generated via axicons are widely used for various applications like optical tweezers or laser microfabrication of transparent materials. The specific intensity profile having high aspect ratio of beam width and length in turn generates high aspect ratio void that resembles a needle. In contrast to commonly generated Bessel beam that has a fixed axial intensity distribution. We present a novel method to engineer an optical needle that can have an arbitrary axial intensity distribution via superposition of different cone angle Bessel beams. We analytically describe spatial spectra of an optical needle having arbitrary axial intensity distribution. We also demonstrate a superposition of independent optical needles and analyze the physical limitations to observe well separated optical needles as they are influence by mutual interference of the individual beams. In order to verify our theoretical and numerical results we generate controllable spatial arrays of individual beams with various numbers and spatial separations by altering a spectrum of incoming laser beam via spatial light modulator. Lastly, we numerically examine distortions caused by propagation through planar air-dielectric interface and show compensation method by appropriately modifying spectral masks.

## 1. Introduction

In many applications of laser microfabrication and optical trapping it is advantageous to use laser beams with long depth of focus and narrow transverse intensity distribution [1]. One example of such optical field is a nondiffracting Bessel beam, which exhibits such features in the so-called Bessel zone [2]. These beams are usually generated by axicons and are widely used in such applications as laser micromachining [4-6] and optical tweezers [7]. However due to fixed bell shaped axial intensity distribution these beams are only a single type of the optical needle family [3].

In practical applications it is important to eliminate aberrations caused by planar dielectric material interface (e.g. focusing from air into the volume of bulk material) [8]. Thus, a numerical investigation of the problem with demonstration how the aberration may be eliminated is of a practical importance.

It was shown in Ref. [9] that the superposition of zeroorder Bessel beams with specific axicon angles and complex amplitudes can be used to create a predefined axial intensity distribution which is more practical than axial intensity pattern obtained by conventional conical lens. In this work we introduce a methodology of producing parallel Bessel-like optical needles with controllable individual axial intensity pattern. We present also an experimental implementation of such beams using a spatial light modulator. Lastly we dive into the problem of the aberrations introduced by a planar interface and show, how one could compensate axial intensity distortions due to focusing through air-dielectric interface. The elimination spatial aberrations caused by planar dielectric material interface (e.g. focusing from air into the bulk material) are very important for practical applications as it not only impacts the laser energy deposition efficiency [8] but also the axial intensity distribution of the optical needle. Thus, a numerical investigation of the problem with demonstration how the aberration may be eliminated is of a practical importance.

## 2. Axial intensity control in an optical needle and translation of optical needles

In this section we present the theoretical basis for creation of arrays of parallel Bessel-like optical needle beams with controlled axial intensity pattern. Here we use ideal Bessel beams as a basis functions. They are obtained from an angular spectrum described by a Dirac's delta function [10]. These nondiffracting beams are well enough approximations of experimentally observed intensity distributions near the focal point of the Fourier lens.

### 2.1 Axial intensity control in an optical needle

Ideal nondiffracting Bessel beam is a solution of scalar Helmholtz equation in circular cylinder coordinates

$$\psi(\rho,\phi,z) = J_m(k_\rho\rho)\exp(im\phi + izk_z), \qquad (1)$$

where $\psi$ - is electric field, $J_m$ - the $m$ - th order Bessel function, $\rho,\phi,z$ - cylindrical coordinates, $k_\rho$, $k_z$ - radial and axial wavenumbers respectively and $m$ is a topological charge of the Bessel beam [10].

Any solution of scalar Helmholtz equation can be represented as a 3D integral containing plane waves with different wave vectors, which can be further reduced to a 2D integral, if the fields are axisymmetric, see [10]. This two-dimensional field representation is based on Fourier-Bessel transform and can be rewritten for our purposes by changing the integration variables from $k_\rho$ to $k_z$. The axial component of the wave vector will enter into the integral as a Fourier transform.

This approach to the engineering of axial profiles is discussed in great detail in Ref [9]. By enforcing the radial coordinate to be zero, one ends up in a Fourier series (with respect to axial coordinate $z$) using a superposition of Bessel beams (1), where a Fourier integral is defined as

$$\Psi(\mathbf{r}) = \int_{-\infty}^{\infty} A(K_z + k_{z0})\psi(\mathbf{r}; K_z) dK_z . \quad (2)$$

We assume, that $k_z = k_{z0} + K_z$, where $k_{z0}$ is a carrier wave vector and $A(k_z)$ - complex amplitude of the spatial spectra (i.e. of the each individual Bessel beam component). The function $\psi$ for the case $m = 0$ has values on axis $\psi = \exp(ik_z z)$, therefore the Eq. (2) on-axis will be an expression for the Fourier spectrum of a selected axial intensity distribution $\Psi(0, z) = f(z)$, and the term $k_{z0}$ is used to shift the spectrum to positive $k_z$ values in order to restrict to forward propagating waves only

$$A(k_z) = \frac{1}{2\pi} \int_{-\infty}^{+\infty} f(z)e^{-ik_z z} dz . \quad (3)$$

Thus, a continuous superposition of Bessel beams $\Psi(\mathbf{r})$ defined in (2) with spatial spectrum (3) can exhibit properties of axial intensity similar to those defined by a function $f(z)$.

### 2.2 Translation of an optical needle

In order to control transverse position of an optical needle we use addition theorem of Bessel beams [10]

$$J_n(k_\rho \rho_2)e^{in\phi_2} = \sum_{m=-\infty}^{\infty} J_m(k_\rho \rho_{12})J_{n+m}(k_\rho \rho_1)e^{im(\phi_1-\phi_{12})} , \quad (4)$$

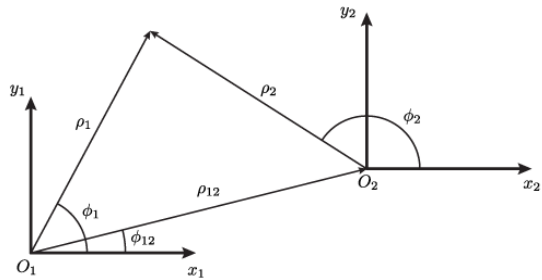

**Fig. 1** Bessel function translation.

where $\rho_1$, $\rho_2$ - cylindrical coordinates of first and second coordinate systems respectively and $\varphi_1$ and $\varphi_2$ are azimuthal angles of first and second coordinate systems. The spectrum of individual Bessel beam is Dirac delta function multiplied by azimuthal phase term, therefore the spectrum

of translated Bessel beam is a sum of many terms with Dirac delta functions

$$\hat{\psi}(k_\rho, \phi_k) = \sum_{m=-\infty}^{\infty} J_m(k_\rho \rho_{12})e^{-im\phi_{12}} \frac{i^m e^{im\phi_k} \delta(k\sin\theta - k_\rho)}{k_\rho} . \quad (5)$$

In this way the Bessel beam with origin at shifted point $O_2$ may be expanded as a superposition of Bessel beams in the unshifted origin $O_1$ (see Fig.1).

Let us assume, we would like to have a number $p = 1, 2, \ldots P$ of independent parallel optical needles each with its own axial profile and position $(x_p, y_p)$ in the transverse plane. We use the superposition principle and express the resulting spatial spectrum as a sum

$$\hat{\Psi}(k_x, k_y) = \sum_p \int_{-\infty}^{\infty} A_p(K_z + k_{z0})\hat{\psi}(k_x, k_y; x_p, y_p) dK_z . \quad (6)$$

here $A_p(k_z) = \frac{1}{2\pi} \int_{-\infty}^{\infty} f_p(z)e^{-ik_z z} dz$ is the Fourier transform of the individual axial intensity profile $f_p(z)$, and $\hat{\psi}(k_x, k_y; x_p, y_p)$ is a Bessel beam's spatial spectrum, when it is shifted in the transverse plane to the point $(x_p, y_p)$.

### 3. Compensation of the aberration due to the planar interface

In order to analyze aberrations caused by air-dielectric interface we extend our scalar beam description to a vector one by assuming that the beam is $x$-polarized

$$\mathbf{V}(\theta_1, \phi) = \hat{\mathbf{x}}\hat{\Psi}(\theta_1, \phi) \quad (7)$$

here $\theta_1$ is the incidence angle on the air-dielectric interface, $\hat{\Psi}$ is the angular spectrum as defined in (7). We note also that Bessel beams produced by conventional conical lens do not suffer spherical aberration as the beams contain single angles of incidence. The refraction only changes the Bessel angles but do not create aberrations as for a Gaussian beam. In our case, we have a continuous integral of individual Bessel beams with different angles, so the resulting beam experiences aberrations due to the planar interface.

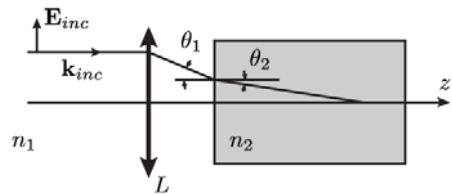

**Fig. 2** Principal scheme of focusing trough air-dielectric interface. Here $L$ – focusing lens, $n_1, n_2$ refractive indices, $\mathbf{E}_{inc}$ - incident field focused by the lens $L$, $\theta_1, \theta_2$ - focusing angles.

According to [8,10] the transmitted field in dielectric medium $\mathbf{E}_t$ can be expressed as the following integral over focusing angles

$$\mathbf{E}_t = \int_0^{2\pi} \int_0^{\theta_{max}} \left\{ t^s(\theta_1)[\mathbf{V}(\theta_1, \phi)\hat{\phi}]\hat{\phi} + t^p(\theta_1)[\mathbf{V}(\theta_1, \phi)\hat{\theta}_1]\hat{\theta}_2 \right\} \times$$
$$\times e^{i(k_x x + k_y y + k_z z)} \sin\theta_2 \sqrt{\cos\theta_2} \, d\theta_2 d\phi, \quad (8)$$

here $t^s$, $t^p$ are Fresnel coefficients [8], $\theta_1$, $\theta_2$, $\phi$ - focusing angles as depicted in Fig. 2

From the analysis of the integral (8) we note that the optical beam in the second medium is distorted due to Snell's law and Fresnel's coefficients. If the ratio $t^p / t^s$ of Fresnel's coefficients is close to the unity then we can correct distortions just by accounting for a change in angles due to Snell's law. This correction is done by substitution $\theta_1 \mapsto \theta_2(\theta_1)$ in the focusing integral kernel (8)

$$\mathbf{E}_t = \int\limits_0^{2\pi} \int\limits_0^{\theta_{max}} \left\{ t^s(\theta_1)[\mathbf{V}(\theta_2,\phi)\hat{\phi}]\hat{\phi} + t^p(\theta_1)[\mathbf{V}(\theta_2,\phi)\hat{\theta}_2]\hat{\theta}_2 \right\} \times$$
$$\times e^{i(k_x x + k_y y + k_z z)} \sin\theta_2 \sqrt{\cos\theta_2} \, d\theta_2 \, d\phi, \qquad (9)$$

such change corresponds to the change in the spatial spectrum.

To illustrate this method of distortions compensation let us analyze the focusing of a circle of parallel optical needles (see Fig 3) from air into dielectric medium with refractive index $n_2 = 1.5$ see first column or index 1, and without, see second column or index 2. Here (a$_{1,2}$, b$_{1,2}$) – are amplitude and phase distributions of spatial spectra and c$_{1,2}$ are electric field intensity distributions. The planar air-dielectric medium interface is located at $z = 0$, the air is the region with $z < 0$.

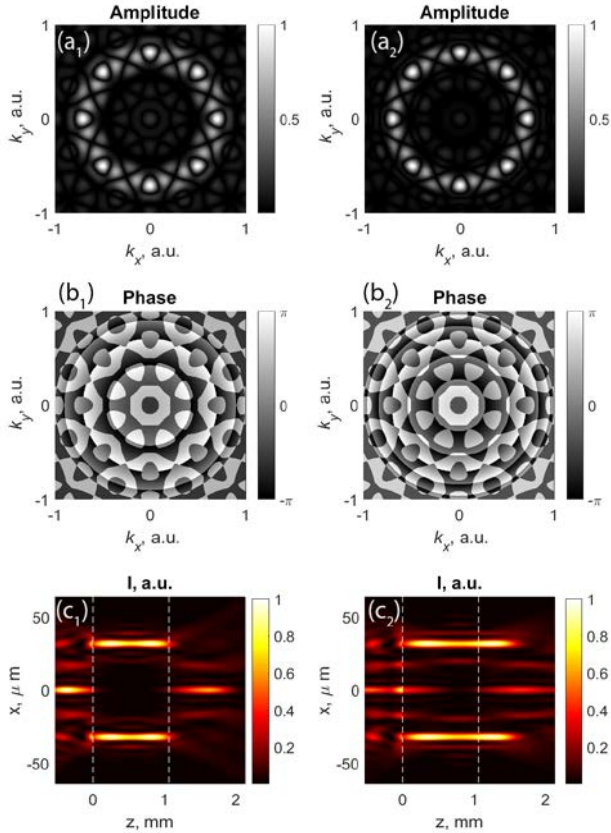

**Fig. 3** Comparison of situations with compensation of distortions due to focusing trough air-dielectric interface ($n_1 = 1$, $n_2 = 1.5$),

Results of our numerical simulations are presented in Fig. 3. First of all, we observe that the presence of a planar interface introduces two types of distortions (see the second column or the right side): the longitudinal one, due to the change in the physical angles and a weaker transverse one, which can be noticed upon careful comparison of the beam profiles. The spatial spectra with angles adjusted so, that upon the entrance through the planar interface, the Snell's law enforces proper axial shape of individual needles, see first column (or the left side) in Fig. 1.

## 4. Experimental setup

Verification of numerically simulated arrays of optical needles in the air were performed using a phase-only spatial light modulator (SLM) together with an optical set-up depicted schematically in Fig. 4. The linearly polarized beam of the wavelength of 532 nm were used for the experiments. The beam is limited to only a small central part (8 mm diam.) of its expanded diameter using a diaphragm to achieve a more uniform intensity over the matrix of the SLM. The beam splitter (BS) cube ensures that the incident angle of zero degree is achieved. A reflected beam has a phase, which was modified via phase delays induced by the SLM. The reflected beam undergoes a rescaling inside a 4f imaging system and is further Fourier transformed by a Fourier lens. Axial and transverse intensity profiles are captured with an optical imaging system mounted on a linear translation stage. We note that we achieve a x27.4 transverse magnification and approximately a x6 longitudinal magnification in our setup, when compared to numerically simulated beams.

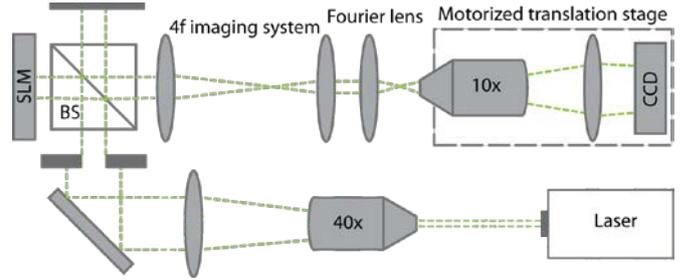

**Fig. 4** Optical set up of the experiment. CW laser, lenses (4f imaging and Fourier), objectives (40x and 10x), spatial light modulator (PLUTOVIS-006-A, HOLOEYE Photonics AG), beam splitter BS and CCD camera.

## 5. Results and discussion

Experiments were carried out using a SLM to verify diffraction of such beams in the air and to examine our control capabilities while controlling a) individual needles and b) their arrays, as well as c) distortions caused by destructive interference between the individual needles inside an array.

### 5.1 Axial intensity profile of single needle

Here we aim to create a constant step-like axial profile. For our experiments we select the function $f(z)$ of the axial profile to be a super-Gaussian function

$$f(z) = \exp\left[-\left(\frac{z - z_0}{z_0}\right)^{2N}\right], \qquad (10)$$

where $z_0 = L / [2\log(2 / 2^{1/N})]$ is a parameter controlling the axial intensity full width at half maximum (FWHM), $N$ - is the order of super-Gaussian function, $L$ is the length at FWHM.

This choice enables for smoother axial intensity profiles (fluctuations around the desired step-like due to the Gibbs phenomenon are smaller both in the numerical simulations and in the experiment).

Firstly, we examine experimentally axial intensity profiles described by the Eq. (10). The main aim here is to achieve a smooth intensity profile over length $L$ with steep edges at the beginning and the end. The shape of this function is strongly dependent on the order $N$ of the super-Gaussian function. For low $N$ values ($N$=1-4) the edge steepness is poor, but the axial profile itself is rather smooth. When increasing the number $N$, we increase also the edge steepness but we lose the smoothness.

Experimental results for a single case of an optical needle with $L$=1 mm are depicted in Fig. 5 which are the best results ($N = 7$) which we have achieved. In this case we observe the most optimal balance between two competing factors, thus, giving us a steep and even axial intensity profile. We will further use super-Gaussian axial profiles with $N = 7$.

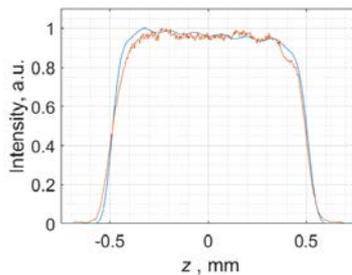

**Fig. 5** A comparison of experimentally measured (orange line) and numerically calculated (blue line) super-Gaussian ($N = 7$) axial intensity profiles of an optical needle of the length $L = 1$ mm generated with our optical set-up.

### 5.2 Arrays of optical needles

Next, we are using in further experiments the same beam parameters as in previous section. Here, we compare experimentally and numerically obtained transverse profiles of three parallel optical needles positioned in one row with spatial separation of $\rho_{12} = 60 \lambda$, see Fig. 1. We observe a good agreement between experimental results and numerical simulation as the direct comparison shows only minor differences in transverse intensity profiles (Fig. 6), which indicates both proper work of our experimental set-up and the correctness of theoretical methods. Some interference of nearby beams occurs in both cases, which can be seen as distortion of typical ring system around Bessel-like beams. All transverse profiles are captured at the middle of the axial profile which is set to be at the Fourier plane.

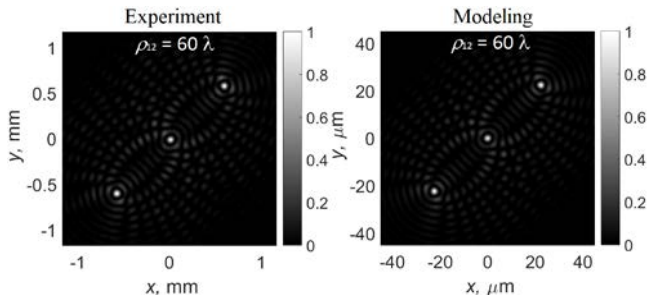

**Fig. 6** Intensity distribution of experimentally obtained and numerical calculated parallel arrays of three "optical needles". The axial length $L$ of individual needle is $L = 1$mm, $N = 7$, the spatial separation $\rho_{12}$ between the needles in the transverse plane is $\rho_{12} = 60 \lambda$.

Spatial separation has a big influence on the formation of individual optical needles. As we try to bring them closer, by lowering the individual separation $\rho_{12}$, the destructive interference tends to increase and optical needles are no longer generated correctly. On the other hand, by increasing the separation distance $\rho_{12}$, distant needles lose some intensity due to limitations of our set-up, see Fig. 7, which also limits the positioning capabilities of the needles using the method described here.

Length of the array is also an important factor to consider as it also heavily affects the occurrence of destructive interference. As it can be seen from Fig.8, an array of individual needles with shorter individual lengths $L$ shows nearly no interference between individual parts. As one could expect, longer individual optical needles in the array cause more destructive interference. This can be understood considering the fact, that the creation of individual Bessel zone requires some volume for the plane waves lying on the cone to interfere and create an optical needle. The longer the needle the larger is the volume of this effective Bessel zone.

Furthermore, we demonstrate an ability to control the length of an individual optical needle inside the array separately. Here, we construct an array of four optical needles with different individual lengths (Fig.9). The influence of the individual length $L$ can be also noticed in this experiment, as the interference between adjacent elements increasingly gets stronger as adjacent optical needles gets longer (Fig. 9 b)).

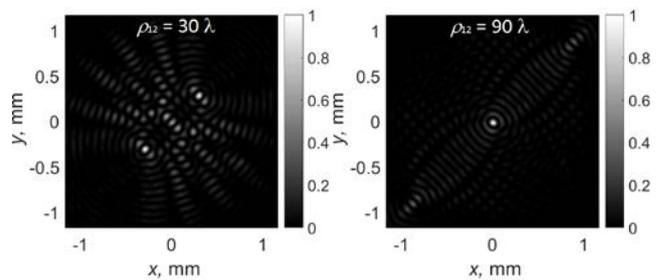

**Fig. 7** Intensity distribution of experimentally obtained parallel arrays of three "optical needles". The axial length $L$ of individual needle is $L = 1$mm, $N = 7$, the spatial separation $\rho_{12}$ between the needles in the transverse plane is depicted on the graph.

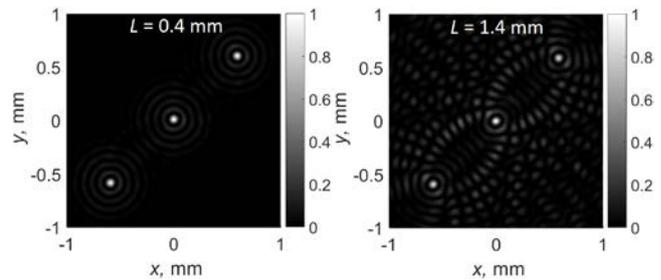

**Fig. 8** Intensity distribution of experimentally obtained parallel arrays of three "optical needles". The axial length $L$ of individual needle is $L = 0.4$ mm and 1.4 mm, $N = 7$, the spatial separation $\rho_{12}$ between the needles in the transverse plane is $\rho_{12} = 60 \lambda$.

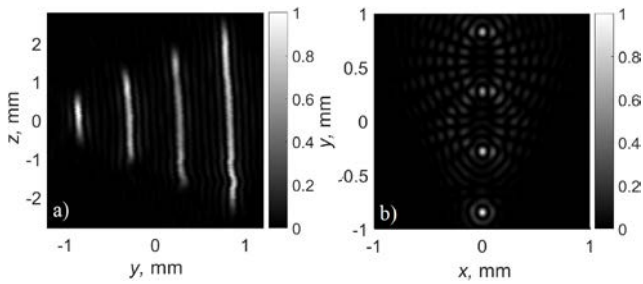

**Fig. 9** Intensity distribution of experimentally obtained parallel arrays of four "optical needles". Intensity profiles are depicted in a) *yz* and b) *xy* planes. The individual lengths L in the array are *L*= 0.2 mm, 0.4 mm, 0.6 mm and 0.8 mm.

Finally, we present here our capability to form complex spatial structures by placing individual optical needles in specific places in the focal plane. A good example would be a situation when we form a circle of eight separate needles of individual lengths $L = 0.5$ mm (Fig. 10). Shorter beams and right spatial separation between allows us to minimize the destructive interference and form a structure with clearly expressed individual needles.

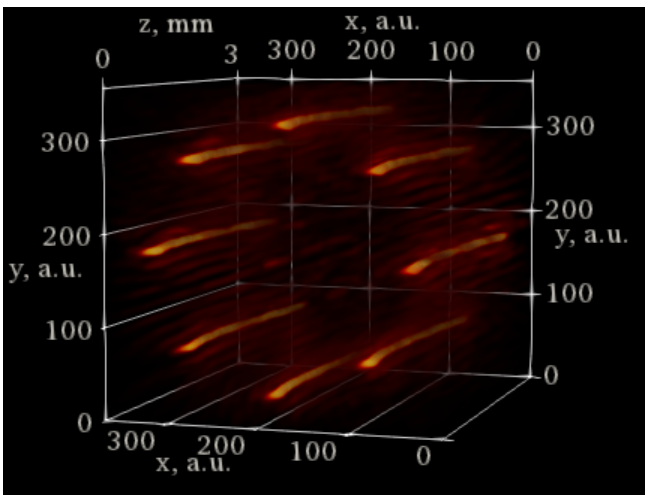

**Fig. 10** Three dimensional depiction of the experimentally measured intensity distribution in the array of eight optical needles with individual length $L = 0.5$ mm. Optical needles are position so that they create a circle.

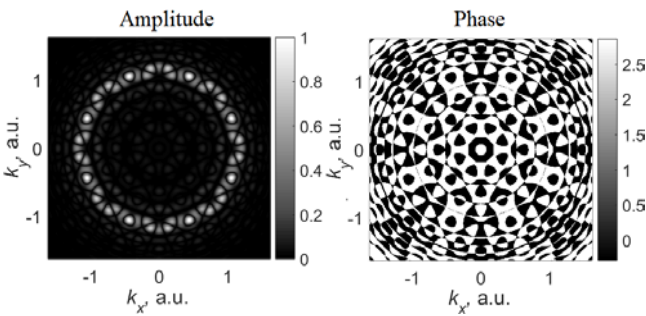

**Fig. 11** Amplitude and phase distributions of the spatial spectra of a circular array containing eight optical needles, depicted in the Fig. 10.

The spatial spectra of such structure is depicted in Fig. 11. We see from the amplitude distribution of the spatial spectra that it contains not only a single ring (as expected for Bessel-Gaussian beam) but a few more, not so intense rings which appear as we superpose multiple beams to control the axial profile. The rings also contains some intensity modulation which occurs as we create and superpose translated beams (see, Eq. (7)).

The phase structure of the spatial spectra is here rather complicated, what can be seen as a manifestation of the complexity of this rather simple structure while reproducing it with plane waves.

## 6. Conclusion

We have presented a flexible technique, which enables us to create experimentally controlled arrays of parallel optical needles with independent axial intensity profiles. We have analyzed how the separation between individual optical needles interplays with the individual lengths of the optical needles. We show, that the destructive interference between adjacent needles is less pronounced when they are of different lengths. Our preliminary analysis shows, that this is caused by the fact, that optical needles of different lengths have different spatial modulation in the Fourier space. The distortion between the neighboring optical needles appears due to the spatial overlapping of the beams. Therefore to achieve the best results it is advisable to separate the beams so that they have limited overlap. The downside of this technique is limiting the smallest separation between the beams, or limits the needle length, or limits the beam width.

Additionally we have introduced a step-like axial intensity profile described using a super-Gaussian function. This has enabled us to avoid problems caused by the Gibb's phenomenon, when the edges of a step function exhibit very sharp oscillations. We have found optimal parameter of the super Gaussian function $N=7$, which ensures both nice smoothness of the profile and sharpness of the edges.

Moreover, we have demonstrated the proof-of-concept implementation of the technique, which allows for compensation of various distortions due to aberrations introduced by a planar interface between air and dielectric.

In conclusion, the method presented here allows creation of various spatial intensity distributions in 3D which might be applicable for possible specific microfabrication tasks or optical tweezing set-ups. The implementation of this approach into high power laser systems will enable us to move away from the spatial light modulator, as it cannot sustain high laser powers. However the spatial light modulator can be a versatile device to prototype the geometrical phase element, which are known for their sustainability to high laser powers.

## Acknowledgments and Appendixes

This research is/was funded by the European Social Fund according to the activity 'Improvement of researchers' qualification by implementing world-class R&D projects' of Measure No. 09.3.3-LMT-K-712.